\documentclass[lettersize,journal]{IEEEtran}
\usepackage{amsmath,amsfonts}
\usepackage{algorithmic}
\usepackage{algorithm}
\usepackage{array}
\usepackage[caption=false,font=normalsize,labelfont=sf,textfont=sf]{subfig}
\usepackage{textcomp}
\usepackage{stfloats}
\usepackage{url}
\usepackage{verbatim}
\usepackage{graphicx}
\usepackage{cite}

\usepackage{booktabs}
\usepackage{multirow}
\usepackage{hyperref}
\usepackage{caption}

\begin{document}

\title{LiVeAction: a Lightweight, Versatile, and Asymmetric Neural Codec Design for Real-time Operation}

\author{
Dan Jacobellis and Neeraja J. Yadwadkar \\
University of Texas at Austin \\
\texttt{danjacobellis@utexas.edu, neeraja@austin.utexas.edu}
}

\maketitle

\begin{abstract}
Modern sensors generate rich, high-fidelity data, yet applications operating on wearable or remote sensing devices remain constrained by bandwidth and power budgets. Standardized codecs such as JPEG and MPEG achieve efficient trade-offs between bitrate and perceptual quality but are designed for human perception, limiting their applicability to machine-perception tasks and non-traditional modalities such as spatial audio arrays, hyperspectral images, and 3D medical images. General-purpose compression schemes based on scalar quantization or resolution reduction are broadly applicable but fail to exploit inherent signal redundancies, resulting in suboptimal rate-distortion performance. Recent generative neural codecs, or tokenizers, model complex signal dependencies but are often over-parameterized, data-hungry, and modality-specific, making them impractical for resource-constrained environments. We introduce a \underline{Li}ghtweight, \underline{Ve}rsatile, and \underline{A}symmetric neural codec architecture (LiVeAction), that addresses these limitations through two key ideas. (1) To reduce the complexity of the encoder to meet the resource constraints of the execution environments, we impose an FFT-like structure and reduce the overall size and depth of the neural-network-based analysis transform. (2) To allow arbitrary signal modalities and simplify training, we replace adversarial and perceptual losses with a variance-based rate penalty. Our design produces codecs that deliver superior rate-distortion performance compared to state-of-the-art generative tokenizers, while remaining practical for deployment on low-power sensors. We release our code, experiments, and python library at \url{https://github.com/UT-SysML/liveaction}.

\end{abstract}


\section{Introduction}

Modern sensors—from wearables and medical devices to satellites—generate rich streams of high-resolution data~\cite{engel2023project, zhang2022progress}. Efficient compression is critical for applications in health monitoring, remote sensing, and autonomous systems, as these deployments operate under strict power and bandwidth constraints. Standardized codecs (JPEG and MPEG) provide strong bitrate–quality trade-offs at low computational cost, but their human-centric design makes them unsuitable for machine-perception tasks and non-standard modalities where perceptual quality is not the target\cite{jacobellis2025learned}.

General-purpose methods, such as scalar quantization~\cite{davisson1968theoretical} and resolution reduction~\cite{kortman1967redundancy}
remain widely used for their simplicity and universality. They apply to arbitrary signals, provide analytical guarantees on information loss, and combine easily with domain-specific approaches~\cite{khani2021efficient, chen2025estimating}. But, being agnostic to real-world data, they fail to exploit inherent redundancies, leading to poor rate–distortion performance~\cite{davisson1972rate}.

Recent advances in deep neural network (DNN)–based autoencoders~\cite{balle2017end, he2022elic} and generative codecs~\cite{ramesh2021zero, agarwal2025cosmos} show that data-driven models can capture complex signal dependencies, greatly improving compression efficiency and realism. These tokenizer-style codecs use learned transforms and perceptual losses to reconstruct high-quality outputs at low bitrates but remain impractical for resource-constrained settings. Their deep, wide encoders dominate computational cost, and their architectures are often modality-specific.
Additionally, generative codecs often depend on perceptual or adversarial losses tuned to human perception, making them ill-suited for scientific or machine-perception tasks. Such objectives are undefined for many signal types and can destabilize training, preventing these models from serving as general-purpose codecs, especially in low-power or embedded settings.

To address these limitations, we propose LiVeAction, a \underline{Li}ghtweight, \underline{Ve}rsatile, and \underline{A}symmetric neural codec designed to achieve efficient, high-fidelity compression across diverse signal modalities. LiVeAction is built to meet three primary goals: (1) extreme computational encoding efficiency, (2) competitive rate–distortion performance, and (3) versatility across signal modalities.

\begin{figure*}[t]
  \centering
  \includegraphics[width=.9\textwidth]{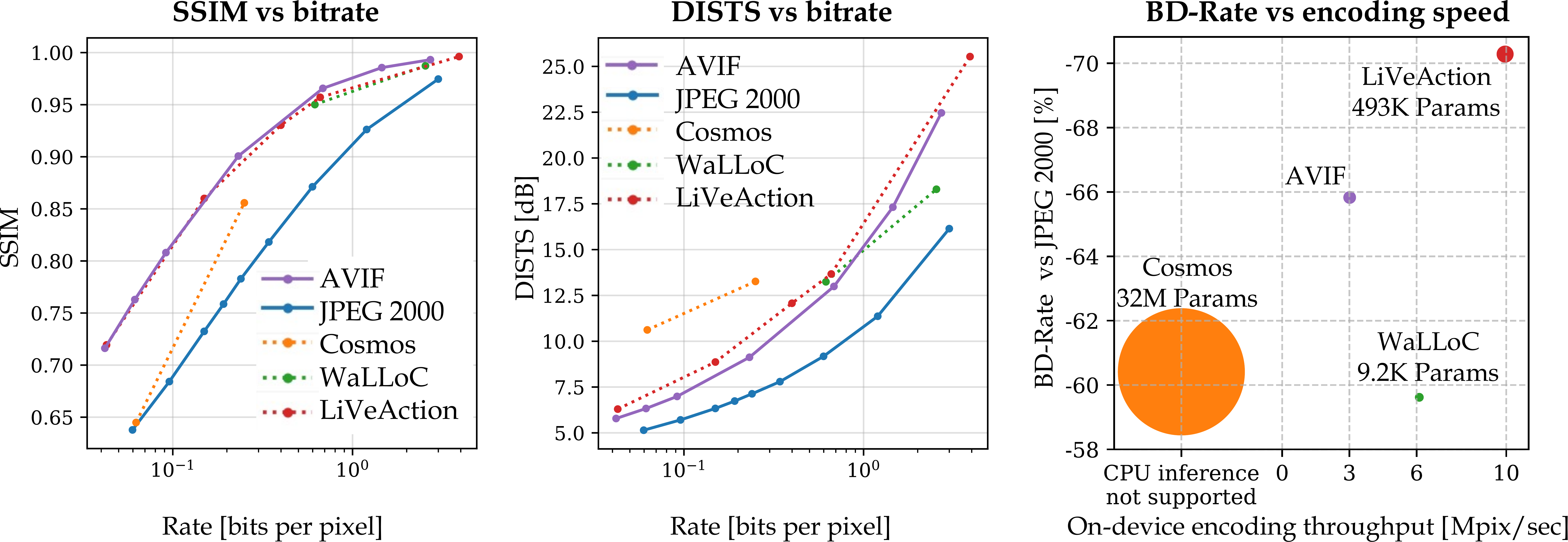}
  \caption{Rate-distortion-complexity trade-off for RGB images measured on the kodak dataset. BD-rate is averaged between SSIM and DISTS~\cite{ding2020image}). Throughput is measured on a low power mobile CPU (Intel Arrow Lake 255U).}
  \label{fig:ratedistortion}
\end{figure*}

\vspace{1mm}
\textbf {Extreme computational encoding efficiency.} 
Real-time sensing on mobile or remote platforms demands encoders that are  computationally efficient and power-conscious. Most neural autoencoders use symmetric architectures, where analysis and synthesis transforms share nearly identical DNN layers~\cite{evans2025stable, agarwal2025cosmos}. However, increasing encoder depth or width yields diminishing returns~\cite{hansen2025learnings}.
LiVeAction adopts an asymmetric design with a lightweight encoder that minimizes computation while preserving representational quality. 
LiVeAction improves efficiency using structured, FFT-inspired operations instead of dense projections. These impose a block-diagonal structure reminiscent of ShuffleNet~\cite{zhang2018shufflenet} and Monarch matrices~\cite{dao2022monarch, fu2023monarch}, allowing multiple layers with alternating nonlinear activations at roughly the cost of one dense layer.

\vspace{1mm}
\textbf{Competitive rate-distortion performance.}
To enable applications with severe bandwidth limitations, the rate-distortion performance must match or exceed conventional standards like JPEG or MPEG. Existing autoencoder designs (e.g. Stable diffusion~\cite{rombach2022high}, Stable Audio~\cite{evans2025stable}, and Cosmos~\cite{agarwal2025cosmos}) rely heavily on perceptual and adversarial losses, enabling the decoder to synthesize realistic, but  hallucinated details.
Prior work shows that removing these losses can improve compressed-domain learning by maximizing the dimension–distortion trade-off~\cite{jacobellis2025learned}.
In LiVeAction, the training objective is purely to optimize the rate-distortion trade-off, similar to learned image compression systems~\cite{balle2017end}. To simplify the training process and increase accessibility for new modalities, we replace the continuously-relaxed probability model and auxiliary optimizer with a simplified rate penalty based on the sample variance. Compared to codecs with generative or adversarial losses, this formulation requires fewer hyperparameters and provides stable training for a wide range of signal types using thousands, rather than millions, of training examples.

\vspace{1mm}
\textbf{Versatility for use with any modality.}
LiVeAction is designed for architectural and loss-function generality to support diverse sensing applications. Prior autoencoders are often tied to specific modalities through custom objectives such as LPIPS~\cite{zhang2018unreasonable}, optical flow loss~\cite{agarwal2025cosmos}, or adversarial losses~\cite{zeghidour2021soundstream, mentzer2020high}. In contrast, LiVeAction shows that a simple mean-squared-error (MSE) based rate–distortion objective suffices across modalities, eliminating the need for perceptual losses.
Existing DNN architecture designs also limit versatility. The convolutional and transformer-based architectures underlying previous autoencoders are meticulously engineered for specific modalities.
LiVeAction’s analysis and synthesis transforms are modality-agnostic and apply to any uniformly grid-sampled signal.
Additionally, simple heuristics are sufficient to choose hyperparameters, avoiding costly searches when adapting to new sensors. Together, these design choices reduce development cost while maintaining strong performance across various modalities.

\vspace{1mm}
\textbf{Contributions.} Using LiVeAction, we create codecs for a wide range of  signal types---spatial audio arrays, hyperspectral images, and 3D medical CT---as well as standard audio, image and video signals.
Even compared to state-of the art neural tokenizers using modality-specific designs and trained with orders of magnitude more data and compute, we show improvements in the rate-distortion-complexity trade-off. For example, compared to Cosmos~\cite{agarwal2025cosmos}, LiVeAction provides a 34\% BD-rate improvement while encoding more than 10$\times$ faster (see Fig. \ref{fig:ratedistortion}).

\vspace{1mm}
\section{Background and related work}
\vspace{1mm}
We build on prior work in (1) high-throughput, training-free lossy compression, (2) autoencoder design for compressed learning and generative modeling, and (3) efficiency optimizations in convolution- and attention-based neural network layers.

\textbf{Computationally efficient lossy compression.}
Transform-based standards such as JPEG and MPEG remain dominant for their strong trade-offs among rate, distortion, and computational cost. They combine energy-compacting transforms with tuned quantization matrices to minimize perceptual distortion for human observers. However, many signals fall outside standard audio, image, or video modalities, where imperceptible details may still matter. In such cases, training-free codecs based on scalar quantization offer high throughput and bounded error~\cite{di2016fast, zhao2021optimizing}. While effective for scientific data, they underutilize inherent signal redundancies, yielding poor rate–distortion performance. For sensors with extreme bandwidth limits, modality-specific specialization becomes necessary, motivating learned codecs trained end-to-end from representative data.

\textbf{Autoencoders for compression and learning.}
End-to-end learned compression using autoencoders has surpassed traditional audio~\cite{zeghidour2021soundstream}, image~\cite{balle2017end, mentzer2020high}, and video~\cite{agustsson2020scale} codecs in rate–distortion performance. Initially, high design and runtime complexity limited adoption, but this changed with the advent of latent generative modeling, where generative dimensionality-reducing autoencoders (GDR-AEs) accelerated high-resolution autoregressive~\cite{ramesh2021zero} and diffusion models~\cite{rombach2022high}. GDR-AEs were later repurposed for discriminative representation learning~\cite{park2023storage, li2023mage} and now underpin state-of-the-art AI models across audio~\cite{copet2024simple, defossez2024moshi}, image~\cite{hurst2024gpt, agarwal2025cosmos}, and video~\cite{polyak2024movie, agarwal2025cosmos} domains. However, runtime efficiency, especially of the encoder, has received little attention, as its cost is overshadowed by the massive models it supports. Improving encoder efficiency is therefore essential for autoencoders that both compress high-resolution data at the edge and accelerate downstream models in the cloud.

\textbf{Network design for efficient representation learning and compression.}
Prior work improved the efficiency of convolutional and attention-based layers used in autoencoding high-resolution signals for both representation learning and compression. ShuffleNet~\cite{zhang2018shufflenet} and Monarch~\cite{dao2022monarch, fu2023monarch} replace standard convolutional and MLP layers with FFT-like structured matrix operations. Squeeze-and-Excitation networks~\cite{hu2018squeeze} introduce lightweight channel attention, while EfficientViT~\cite{cai2023efficientvit} employs ReLU linear attention to scale to high-resolution.
The computational efficiency of compressive autoencoders has since improved dramatically. Finite scalar quantization (FSQ)~\cite{mentzer2024finite} unified earlier designs—vector-quantized VAEs~\cite{van2017neural} and soft-quantized rate–distortion autoencoders~\cite{balle2017end}. Recent models sandwich an FSQ-based bottleneck between invertible operations that trade spatial or temporal resolution for channel capacity. PatchMixer~\cite{muckley2025architecture}, ViTok~\cite{hansen2025learnings}, and DCVC-RT~\cite{jia2025towards} use local patchifying or tubelet embedding, while WaLLoC~\cite{jacobellis2025learned} and Cosmos~\cite{agarwal2025cosmos} employ wavelet packet transforms for additional energy compaction. Despite these advances, current methods still lag standardized codecs in the rate–distortion–complexity trade-off~\cite{jia2025towards}.

\section{Proposed method: design and implementation}

In order to enable applications of machine perception using diverse signal modalities in resource-constrained environments, Live Action is designed around three key goals: (1) extreme runtime computational encoding efficiency (2) competitive rate-distortion performance, and (3) flexibility for use with arbitrary modalities.

\begin{figure*}[t]
    \centering
    \includegraphics[width=.8\linewidth]{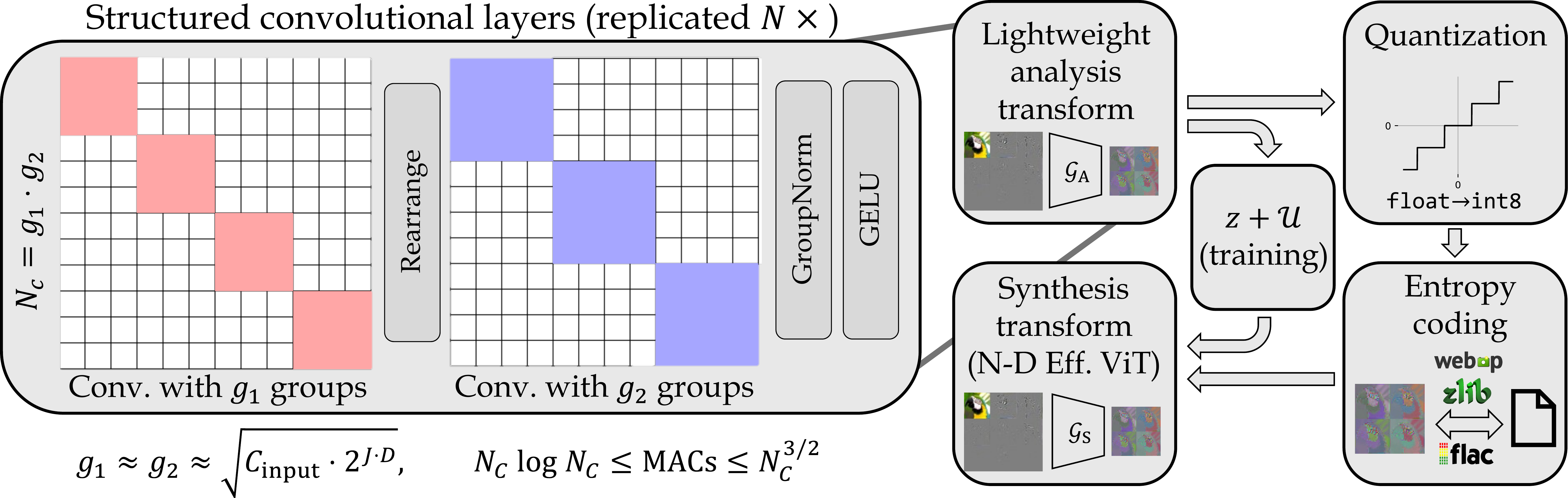}
    \caption{Proposed design. The analysis transform uses a lightweight DNN with block-diagonal structured operations.}
    \label{fig:design}
\end{figure*}

\textbf{Overview and codec workflow.} 
LiVeAction inherits the overall architecture from WaLLoC~\cite{jacobellis2025learned} and Cosmos~\cite{agarwal2025cosmos}, consisting of an FSQ~\cite{mentzer2024finite} based autoencoder sandwiched between the WPT and IWPT. However, our asymmetric design introduces several changes to the DNN-based transforms and training procedures.
Fig. \ref{fig:design} provides an overview of the codec workflow and structured convolution layers, which we describe next.

Let \(x\in\mathbb{R}^{C\times T_1\times\cdots\times T_D}\) signal with \(D\) spatio-temporal dimensions and \(C\) channels. The end-to-end codec is
\begin{equation}
\hat x=\operatorname{IWPT}_{J}\circ
\; C^{-1} \circ
\mathcal G_{\!S}\circ
\Phi^{-1}\circ
\mathcal Q\circ
\Phi\circ
\mathcal G_{\!A}\circ
C \circ
\operatorname{WPT}_{J}(x).
\label{eq:pipeline}
\end{equation}

\noindent \(\operatorname{WPT}_{J}\) and \(\operatorname{IWPT}_{J}\) apply \(J\) dyadic filter bank stages using the Cohen--Daubechies--Feauveau~9/7 filters to trade spatiotemporal resolution for frequency resolution.
The analysis transform \(\mathcal G_{\!A}\). consists of \(d_{\text{enc}}\) factorized group-convolution residual blocks followed by a \(1{\times}1\) projection to latent width \(C_z\).
A factorized convolution replaces a dense kernel by two grouped convolutions with groups \((g_1,g_2)\) chosen to minimize MACs (Monarch/ShuffleNet-style), yielding an FFT-like block-diagonal structure.
GELU is used as the nonlinearity. The group normalization uses 8 groups.
$C$ is an Invertible power-law compander
$
    C(x)=\operatorname{sgn}(x)\bigl[(|x|+\varepsilon)^{\gamma}-\varepsilon^{\gamma}\bigr],
$
where $\gamma{=}0.4,\;\varepsilon{=}0.1$.
\(\Phi\) is a Non-invertible per-channel Laplacian CDF
$
    \Phi(x)=127\,\operatorname{sgn}(x)\!\bigl(1-e^{-|x|/\sigma_c}\bigr),
$
where $\sigma_c>0$ is learned; $\Phi$ ensures latents lie in \([-127,127]\) (strictly less than 8 bits).
\(\mathcal Q\) is Finite scalar quantization trained using a soft-to-hard scheme: for the first 70\% of training, \(\mathcal Q(x)=x+u,~u{\sim}\mathcal U[-\tfrac12,\tfrac12]\); afterwards the encoder is frozen and \(\mathcal Q(x)=\operatorname{round}(x)\). \(\mathcal G_{\!S}\) is the synthesis transform consisting of EfficientViT linear-attention blocks (generalized to 1/2/3-D), with depth \(d_{\text{dec}}\).\\

\begin{figure}[h]
\centering
\includegraphics[width=0.8\columnwidth]{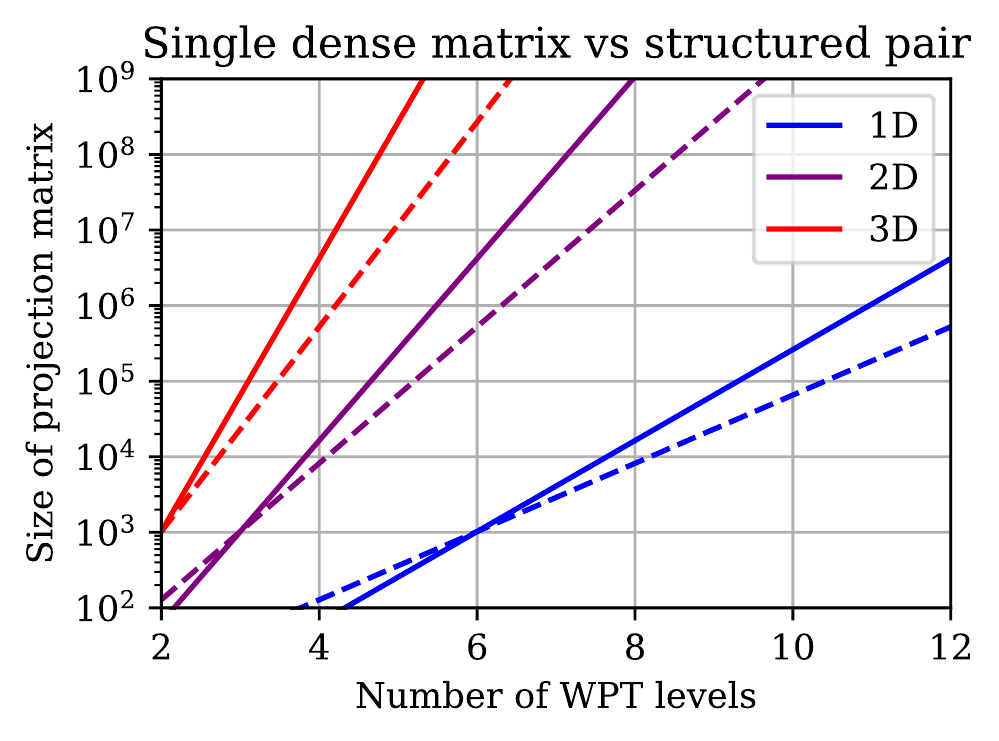}
\caption{Scaling behavior of linear projection (solid line) vs the proposed structured matrix pair (dotted line).}
\label{fig:scaling}
\end{figure}

\vspace{2mm}
\textbf{Lightweight analysis transform for efficient encoding.}
In WaLLoC, the encoder consists solely of a learnable linear projection, trading expressiveness for high efficiency.
Yet, this projection can still be costly.
As an example, consider a spatiotemporal autoencoder for RGB videos.
The WPT maps a $3\times8^3$ RGB video region to 1536 color-frequency bands; projecting these to a $12$-D latent requires a $1536\times12$ matrix-vector product for each local video region.
At 1080p, this results in $> 1.7$ billion FLOPs per second of video for the projection alone.
To significantly increase the computational efficiency of encoding,  LiVeAction replaces this monolithic projection by several grouped convolutional layers, yielding a structured pair with substantially fewer parameters and lower computational requirements compared to a dense matrix, as shown in Figure~\ref{fig:scaling}.
This results in an FFT-like structure for the analysis transform, similar to Shufflenet~\cite{zhang2018shufflenet} and Monarch~\cite{dao2022monarch, fu2023monarch}.
Even using several of these layers with alternating nonlinear activations, added channel attention~\cite{hu2018squeeze}, and group normalization, we achieve encoding throughput competitive with the fully connected linear projection used in WaLLoC.

\vspace{2mm}
\textbf{Linear attention synthesis transform for versatility across modalities.}
The intended applications of LiVeAction---real time sensing on resource-constrained mobile and remote sensors---place extreme demands on the encoder.
However, at runtime, the decoder can be run on powerful cloud GPUs, or even discarded entirely in the case of compressed domain processing. Still, increasing accessibility for new codecs requires high-resolution training to be possible with low or moderate compute resources---not datacenter-scale GPU clusters. Thus, we adopt an  EfficientViT-based design~\cite{cai2023efficientvit}, leading to uncompromised expressiveness while enabling high-resolution training on a single GPU.
We make two modifications to EfficientViT: (1) replacement of batch normalization with group normalization to eliminate differences between train-test behavior~\cite{wu2018group}, and (2) generalization to one and three dimensions to accommodate additional signal modalities other than 2D images.

\textbf{Finite scalar quantization with simplified rate penalty.}
To achieve a high compression rate, we use finite scalar quantization (FSQ)~\cite{mentzer2024finite} a type of learned vector quantization. Unlike standard VQ-AEs~\cite{van2017neural}, which require expensive codebook lookup operations, FSQ uses of a guaranteed dimension bottleneck (typically between 32$\times$ and $128\times$ reduction) combined with scalar quantization to achieve equally efficient coding. Existing FSQ designs typically aim for a small or moderate codebook size (typically $\leq 16$ bits) to support standard cross-entropy losses and increase compression ratio at the cost of objective reconstruction metrics like PSNR. To meet our goal of maximum versatility, we instead opt for much larger codebook size, but include a rate penalty during optimization, similar to the standard approach used in learned image and codecs~\cite{balle2017end, he2022elic, jia2025towards}.
To reduce the design, implementation, and ``operational"~\cite{jia2025towards} complexity, we introduce an extremely simplified formulation for the rate loss.
Assuming that latent activations follow a distribution in the exponential family (e.g. generalized Gaussian), minimizing the rate is equivalent to minimizing the log of the sample variance.
Thus, our overall training objective is to minimize 
\begin{equation}
\mathcal L=\log_{10}\!\bigl\|x-\hat x\bigr\|_{2}^{2}\;+\;\lambda\,
\log_{2}\!\bigl(\operatorname{Var}[\Phi\!\circ\!\mathcal G_{\!A}(x)]\bigr)
\label{eq:loss}
\end{equation}
with a single global hyper-parameter \(\lambda\). The first term is the MSE distortion; the second approximates the latent rate under an exponential-family prior. We set \(\lambda=3\times 10^{-2}\) for all modalities.
Finally, we adopt a soft-then-hard quantization scheme~\cite{guo2021soft}.
During the main training phase, additive noise is used to encourage resilience to quantization~\cite{balle2017end}.
Near the end of training (70 percent in our experiments), the encoder is frozen, and the additive noise is replaced with hard quantization (rounding) for the remainder of the decoder training. After quantization, any entropy coding method can be used, including lossless media codecs (e.g. FLAC, PNG, FFV1, etc) by reshaping the latents to the appropriate dimension. In our experiments, we find that WEBP lossless and JPEG-LS~\cite{weinberger2000loco} provide the best trade-off between compression and computation efficiency for the entropy coding step, though the differences between methods are minor. We include the cost of entropy coding and file storage when measuring throughput.

\textbf{Heuristics for choosing hyperparameter values}. Building a codec using LiVeAction requires choosing hyperparameters. The exact settings used to reproduce our results for each modality are available in the accompanying code repository. Here, we list several heuristics for choosing these hyperparameters for new modalities.

\begin{enumerate}
    \item \textbf{Dimension.} The codec can operate on 1D, 2D, or 3D signals with arbitrary channel count. For many modalities (e.g. single channel audio) the choice of dimension is unambiguous. However, for modalities with high channel count (e.g. the 224 band hyperspectral AVIRIS images), the channels may be treated as an additional dimension. As a rule of thumb, we recommend treating the channels as an additional dimension if both (1) the number of channels is similar to spatiotemporal resolution of the other dimensions and (2) all of the channels have consistent units/scale.
    \item \textbf{Rate-distortion Lagrangian.} In our experiments, all LiVeAction codecs are trained to minimize $\log_{10} \| x-\hat{x}\| + \lambda \log_2(\hat{\sigma})$, with the parameter $\lambda$ controlling the trade-off between rate and distortion. We find that $\lambda=0.03$ provides stable training across all codecs while cutting the average bitrate by about half (about 4 bits per latent channel instead of 8).
    \item \textbf{Latent dimension.} In addition to $\lambda$, the main hyperparameter affecting the compression ratio is the number of latent channels. For natural signals with significant redundancy, we recommend choosing a latent dimension to be $64\times$ lower than the original dimension.
    \item \textbf{Number of levels $J$ in wavelet packet analysis.} With the exception of the projection to and from the latent dimension, all hidden DNN layers operate with a hidden dimension of $C 2^{JD}$, where $C$ is the number of signal channels and $D$ is the dimension. We recommend choosing $J$ such that the hidden dimension is between 512 and 1536.
    \item \textbf{Depth} In our experiments, we find that an encoder depth of 4 and a decoder depth of 8 leads to a good balance between runtime encoding efficiency, decoder training cost, and rate-distortion performance. 
\end{enumerate}

\section{Evaluation}

Using LiVeAction, we train codecs across multiple signal modalities. We next describe the datasets, evaluation metrics, testbed, and baselines used.\\

\textbf{Stereo audio.} We train on the lossless MUSDB18-HQ dataset~\cite{rafii2017musdb18}, progressively raising clip length from 500k (11s) to 2M samples (48s). Training runs for 200k steps (batch size 2). For augmentation, stems (vocals, drums, bass, other) are randomly remixed; evaluation uses the original validation mixes.

\vspace{1mm}

\textbf{Spatial audio.} We train a spatial audio codec for the 7-channel Aria~\cite{engel2023project} microphone array, progressively increasing clip length from 3 to 7 seconds. Training runs for 288k steps with a batch size of 2. 
Evaluation uses the validation split. In addition to PSNR, we measure the signal to spatial distortion ratio (SSDR) and signal to residual distortion ratio (SRDR) to isolate spatial distortion from other impairments~\cite{watcharasupat2024quantifying}.

\vspace{1mm}

\textbf{Image.} The codec is trained on LSDIR~\cite{li2023lsdir}, with resolution increasing from 128$^2$ to 480$^2$ over 500k steps (batch size 16). Evaluation follows~\cite{agarwal2025cosmos} on the 50k-image validation split of ImageNet, resizing all images to height 1024. We also evaluate the rate-distortion performance and top-1 classification accuracy\footnote{Classification accuracy is evaluated on decoded images using the pre-trained \href{https://huggingface.co/timm/eva_giant_patch14_224.clip_ft_in1k}{EVA-CLIP} vision transformer model.} on the ImageNet validation split at low resolution ($224 \times 224$) and on the Kodak dataset.

\vspace{1mm}

\textbf{Hyperspectral.} We extract 1,394 crops (1,300 training, 94 validation; $\sim$0.5 MP each) from 224-band AVIRIS images~\cite{aviris}. The codec is trained for 130k steps with a maximum resolution of $224\times288^2$. Evaluation is performed on full-size images.

\vspace{1mm}

\textbf{3D medical images.} We train a 3D codec on the MEDMNIST 3D dataset~\cite{yang2023medmnist}, with 6 categories of medical volumes: organ, adrenal, fracture, nodule, synapse, and vessel. Resolution increases from $24^3$ to $64^3$ voxels over 863.5k steps.

\vspace{1mm}

\textbf{Video.} We train on 6,000 Vimeo90k~\cite{xue2019video} clips using two 24-frame batches, with resolution increasing from 112$\times$64 to 640$\times$384 over 120k steps. The model is fine-tuned on 3,000 high-resolution Vimeo90k clips (batch size 1), with resolution increasing from 680$\times$384 to 1152$\times$648. Evaluation uses full-length DAVIS~\cite{perazzi2016benchmark} videos at 1920$\times$1080.

\vspace{1mm}

\textbf{Metrics and baselines.} We evaluate the trade off between rate distortion, and complexity~\cite{minnen2023advancing} using compression ratio (CR), PSNR, and per-sample throughput, and report dimensionality reduction (DR) as a proxy for downstream acceleration~\cite{jacobellis2025learned}.
PSNR is computed on signals in $[0,1]$; some works (e.g., Cosmos) use $[-1,1]$, yielding values $6.02$ dB higher ($20\log_{10}(2)$).
We compare with conventional and neural compression systems, including JPEG2000, Stable Audio~\cite{evans2025stable}, EnCodec~\cite{defossez2023high}, Cosmos~\cite{agarwal2025cosmos}, and WalloC~\cite{jacobellis2025learned}.

\subsection{Results and Discussion.}

\begin{figure}[tbp]
\centering
\includegraphics[width=0.8\columnwidth]{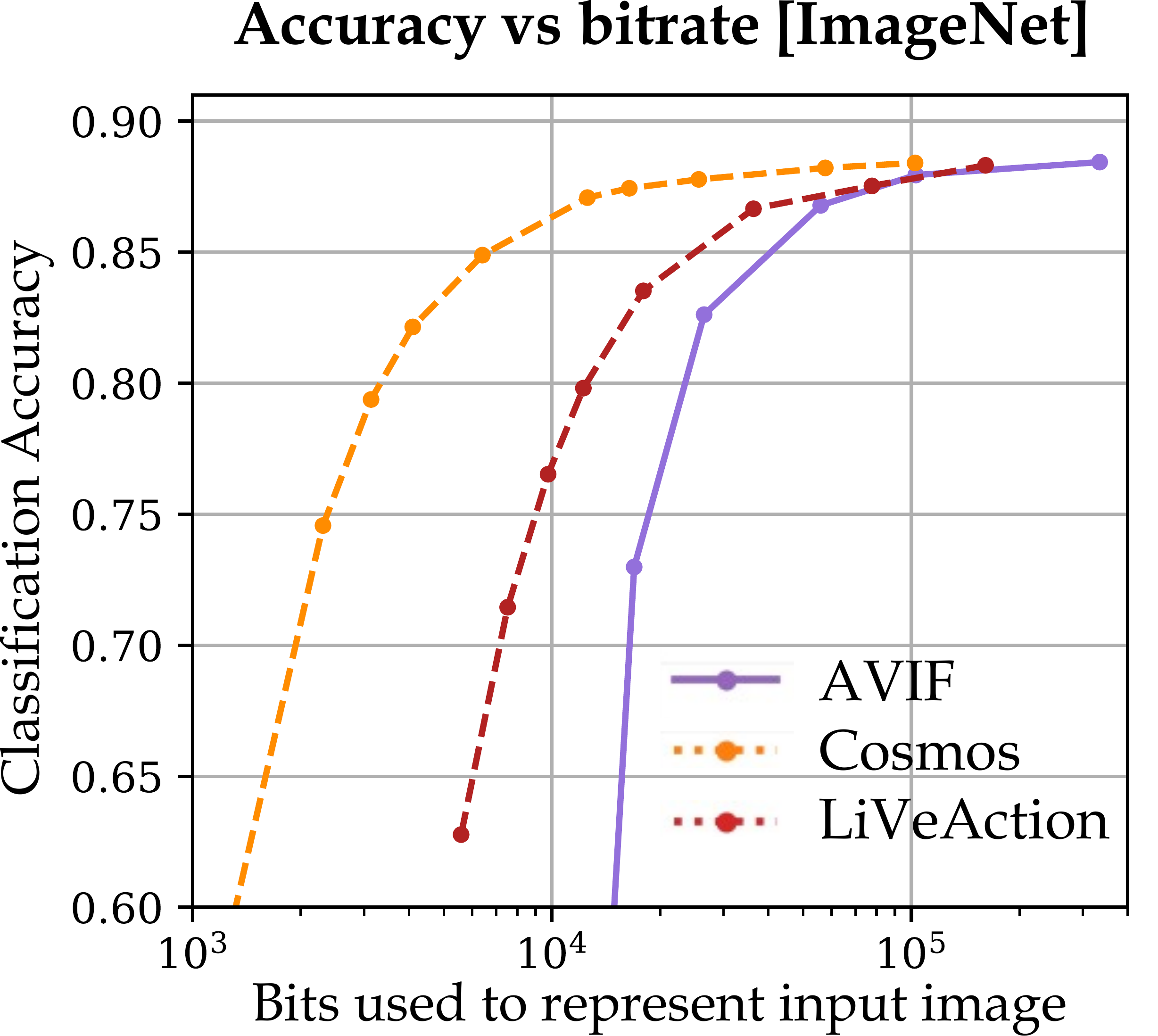}
\caption{Machine perceptual quality of Image codecs measured via Imagenet Classification accuracy. Note: Cosmos is trained on ImageNet, while LiVeAction is not.}
\label{fig:imagenet_acc}
\end{figure}

Tables~\ref{tab:rate-distortion}, \ref{tab:rgb_mobile}, and \ref{tab:cpu_throughput} summarize LiVeAction’s performance across modalities. Figure~\ref{fig:imagenet_acc} shows downstream machine-perception quality for RGB images via ImageNet classification accuracy on decoded outputs. Overall, LiVeAction establishes a superior rate–distortion–complexity frontier, particularly in encoding efficiency on resource-constrained hardware. Despite using simpler training objectives, smaller datasets, and far fewer GPU hours than prior generative tokenizers, LiVeAction remains highly competitive in rate–distortion performance while enabling practical deployment on low-power sensor devices.\\

\begin{table}[htbp]
    \centering
    % \scriptsize
    % \footnotesize
    \begin{tabular}{l l c c c c}
    \toprule
    \multirow{3}{*}{\textbf{Modality}} & \multirow{3}{*}{\textbf{Codec}} & \multicolumn{4}{c}{\textbf{Metrics}} \\
    \cmidrule(l){3-6}
        & & \textbf{DR} & \textbf{CR} & \textbf{Enc.} & \textbf{PSNR} \\
    \midrule
    \multirow{2}{*}{Stereo music}
        & Stable Audio & 64 & 64.0 & 12.31 & 28.42 \\
        & LiVeAction  & 64 & \textbf{195} & \textbf{199.3} & \textbf{36.57} \\
    \midrule
    \multirow{2}{*}{Spatial audio}
        & EnCodec & 5 & 455 & 10.23 & 27.96 \\
        & LiVeAction  & \textbf{64} & \textbf{1013} & \textbf{363.2} & \textbf{33.12} \\
    \midrule
    \multirow{4}{*}{RGB Image}
        & Cosmos DI8 & 32 & 96.0 & 54.96 & 31.20 \\
        & Cosmos DI16 & \textbf{128} & \textbf{384} & 116.5 & 25.08 \\
        & LiVeAction f16c48 & 16 & 34.3 & 58.88 & \textbf{37.81} \\
        & LiVeAction f16c12 & 64 & 140 & \textbf{143.3} & 31.09 \\
    \midrule
    \multirow{2}{*}{Hyperspectral}
        & JPEG~2000 & 1 & \textbf{575} & 12.47$^{\dagger}$ & 18.18 \\
        & LiVeAction & \textbf{64} & \textbf{575} & \textbf{600.1} & \textbf{18.52} \\
    \midrule
    \multirow{2}{*}{3D Medical}
        & JPEG~2000 & 1 & 95.62& 13.60$^{\dagger}$ & 22.00 \\
        & LiVeAction & \textbf{64} & \textbf{209} & \textbf{54.08} & \textbf{24.74} \\
    \midrule
    \multirow{4}{*}{Video}
        & Cosmos DV$4\times8$ & 128 & 96.0$^\ast$ & 7.656$^\ast$ & 28.96 \\
        & Cosmos DV$8\times8$ & \textbf{256} & 192$^\ast$ & 13.73$^\ast$ & 27.43 \\
        & LiVeAction f8c48 & 32 & 79.6 & 33.61 & \textbf{30.24} \\
        & LiVeAction f8c12 & 128 & \textbf{331} & \textbf{52.94} & 27.60 \\
    \bottomrule
    \end{tabular}
    \caption{Rate-distortion-complexity trade-off for each modality. DR is the degree of  dimensionality reduction of. CR is the compression ratio. Encoding throughput (Enc.) is measured in megasamples per second (audio), megapixels per second (images), megavoxels per second (hyperspectral) and frames per second (video). The analysis transform is run on GPU (RTX 4090) and the entropy coding is run on CPU (Intel i9 13900k) with the exception of JPEG 2000$\dagger$, where no GPU acceleration is available. 
    \\$^\ast$ Encoding the entire video in one pass using Cosmos is not possible due to memory constraints. Instead, we encode chunks of 24 frames with 50 percent overlap, resulting in reduced compression ratio and throughput. If no memory constraints were imposed, the CR would be increased by $4\times$ and the throughput would be increased by $3\times$.}
    \label{tab:rate-distortion}
\end{table}

\begin{table}[tbp]
\centering
\begin{tabular}{lcccc}
\toprule
\textbf{Codec}     & \textbf{BD-rate} & \textbf{BD-rate} & \textbf{BD-rate} & \textbf{Throughput} \\
& \textbf{(PSNR)}  & \textbf{(DISTS)} & \textbf{(SSIM)}  & \textbf{(MPix/s)}   \\
\midrule
Cosmos             & +49.61             & \textbf{$-$90.88}    & $-$29.94            & N/A                 \\
WaLLoC             & $-$27.61           & $-$61.71             & $-$57.52            & 6.12                \\
LiVeAction         & $-$36.55           & $-$70.27             & \textbf{$-$70.30}   & \textbf{9.95}       \\
AVIF               &\textbf{$-$64.03}   & $-$60.56             & \textbf{$-$71.10}   & 3.01                \\
\bottomrule
\end{tabular}
\caption{BD-rate relative to JPEG 2000 and encoding throughput on low-power mobile CPU (Intel Arrow Lake 255U) for RGB images. All metrics are measured on the Kodak dataset except for Accuracy, which is measured on ImageNet. Lower BD-rate is better for all metrics.}
\label{tab:rgb_mobile}
\end{table}

\begin{table}[tbp]
    \centering
    \begin{tabular}{l l c c}
    \toprule
    \multirow{2}{*}{Modality} & \multirow{2}{*}{Codec} & \multicolumn{2}{c}{Throughput} \\
    \cmidrule(lr){3-4}
     & & Small Input & Large Input \\
    \midrule
    \multirow{2}{*}{\centering Music (stereo)} & Stable Audio & 88.73 KSamp/s & 229.4 KSamp/s \\
                                               & LiVeAction   & \textbf{323.76 KSamp/s} & \textbf{5012. KSamp/s} \\
    \midrule
    \multirow{3}{*}{\centering RGB Image} & JPEG 2000   & \textbf{6.097 Mpix/s} & 6.333 Mpix/s \\
                                          & Ballé18~\cite{balle2018variational} & 3.440 Mpix/s & 5.106 Mpix/s \\
                                          & LiVeAction     & 5.252 Mpix/s & \textbf{12.28 Mpix/s} \\
    \midrule
    \multirow{2}{*}{\centering Hyperspectral} & JPEG 2000   & 6.298 Mvox/s & 6.448 Mvox/s \\
                                              & LiVeAction  & \textbf{13.76 Mvox/s} & \textbf{14.93 Mvox/s} \\
    \midrule
    \multirow{1}{*}{\centering Video}        & LiVeAction           & 107.6 fps & 2.386 fps \\
    \bottomrule
    \end{tabular}

    \caption{Encoding throughput on high-power CPU (Intel Raptor Lake i9-13900k). Cosmos models are not supported for CPU inference. Sizes for small and large inputs are $2^{12}$ samples (85 ms) and $2^{16}$ samples (1.3 s) [music]; 240p and 1080p [images]; $224^3$ voxels and $224\times1024^2$ voxels [hyperspectral]; 240p and 1080p [video]. }
    \label{tab:cpu_throughput}
\end{table}

\textbf{Music (stereo).} The VAE in Stable Audio generates Gaussian but high-entropy latents, requiring fp16 precision to avoid artifacts. In contrast, LiVeAction’s FSQ design with a rate penalty produces lower-entropy latents, achieving 3$\times$ higher compression. Stable Audio’s reliance on perceptual and adversarial losses causes cross-channel inconsistencies, whereas LiVeAction’s MSE loss yields better stereo fidelity and +8 dB PSNR. Its structured encoder operations are also far cheaper than Stable Audio’s CNN layers, providing over 16$\times$ higher throughput.

\vspace{1mm}

\textbf{Spatial audio.} LiVeAction outperforms EnCodec with 12.8$\times$ greater dimensionality reduction ($64\times$ vs $5\times$), 2.2$\times$ higher compression, and 35.6$\times$ faster encoding while improving all distortion metrics---achieving +6.09 dB SSDR and +13.55 dB SRDR.

\vspace{1mm}

\textbf{RGB image.} On low-power mobile CPU (Intel Arrow Lake 255U), LiVeAction achieves the highest encoding throughput (9.95 MPix/s) and strong BD-rate savings relative to JPEG 2000 (-36.55\% PSNR, -70.30\% SSIM, -70.27\% DISTS); Cosmos is not supported on this platform. Compared to prior neural tokenizers, LiVeAction provides comparable reconstruction quality at similar or higher compression ratios while enabling far greater encoding speed. Notably, despite not being trained on ImageNet (unlike Cosmos), LiVeAction matches Cosmos' downstream ImageNet top-1 classification accuracy on decoded images while using 48\% lower bitrate (Figure~\ref{fig:imagenet_acc}).

\vspace{1mm}

\textbf{Hyperspectral.} Compared to JPEG 2000, LiVeAction reduces latent dimensionality by $64\times$ to accelerate downstream models while slightly improving rate–distortion performance. Its DNN-based design also benefits from GPU acceleration, delivering $\sim$70$\times$ higher throughput than CPU-only JPEG 2000 and over 2$\times$ faster encoding even on the same CPU.

\vspace{1mm}

\textbf{3D medical images.} On MEDMNIST 3D, LiVeAction surpasses JPEG 2000 across all metrics, achieving 64$\times$ dimensionality reduction, 2.1$\times$ higher compression, and 2.7dB higher PSNR for improved rate–distortion performance.

\vspace{1mm}

\textbf{Video.} LiVeAction's lightweight encoder design enables single-pass encoding of full-length 1080p videos on a single RTX 4090, avoiding the memory-intensive chunking required by Cosmos. At comparable quality, LiVeAction achieves $>1.7\times$ higher compression ratio (330.7$\times$ vs. 192$\times$) and $>3.8\times$ higher GPU throughput (52.94 fps vs. 13.73 fps). Real-time encoding ($>60$fps) is possible on CPU at low or moderate resolution.

\subsection{Additional experiments.}

\textbf{Ablation of simplified rate loss.} To isolate the effect of the simplified rate loss, we retrained the RGB codec using an explicit rate term. The implementation uses the \texttt{EntropyBottleneck} module from CompressAI\cite{begaint2020compressai} with an auxiliary optimizer.
~\footnote{\href{https://interdigitalinc.github.io/CompressAI/tutorials/tutorial_custom.html}{%
    \url{https://interdigitalinc.github.io/CompressAI/tutorials/tutorial_custom.html}}}.
Results are shown in Table~\ref{tab:ablation-rate-loss}. The approximate rate model provides a 22 percent reduction in bitrate with minor quality impact.

\begin{table}[htbp]
\centering
\begin{tabular}{lccccc}
\toprule
Objective & bpp & PSNR & LPIPS (dB)\\
\midrule
Approximate rate: $\log_2(\hat{\sigma})$ & 0.6456 & 30.8464 & 6.7503 \\
Exact rate + density model & 0.8334 & 31.1914 & 6.8621 \\
\bottomrule
\end{tabular}
\caption{Ablation of the simplified rate loss. The reported bitrate is the actual number of bits after entropy coding, not the rate estimated from the distribution.}
\label{tab:ablation-rate-loss}
\end{table}

\begin{figure*}[htbp]
  \centering
  \includegraphics[width=.9\textwidth]{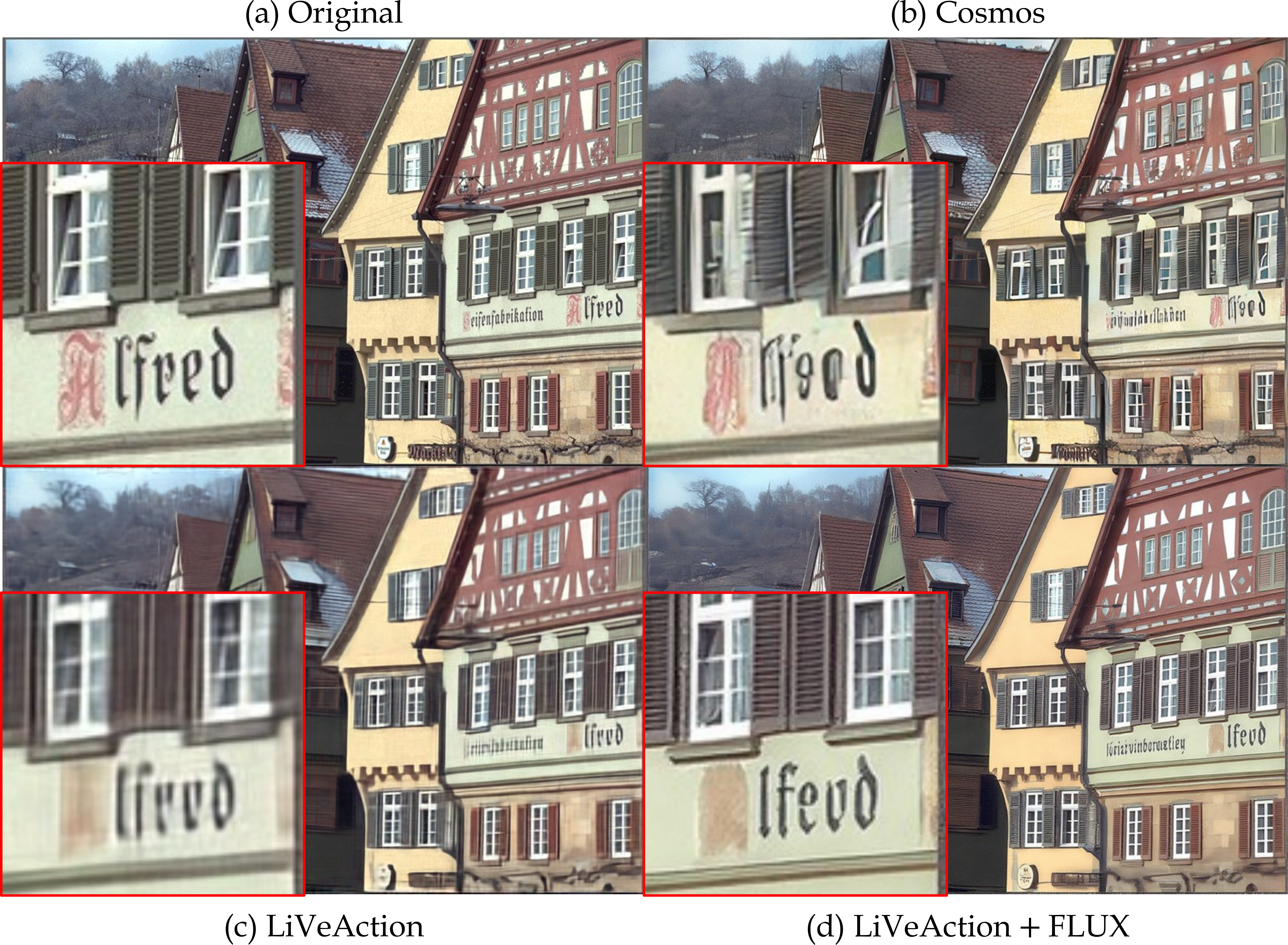}
  \caption{Comparison of Cosmos, LiVeAction, and LiVeAction enhanced using a generative model. Best viewed zoomed in. The image was rescaled before compressing with cosmos to match the rate of two codecs (0.15 bpp).}
  \label{fig:gen_decoder}
\end{figure*}

\textbf{Perceptual quality enhancement using score-based generative model.}
Since LiVeAction omits adversarial and perceptual losses, its decoder does not resynthesize high-frequency details. We show that a separate score-based generative model can enhance perceptual quality post-decoding. Specifically, we use a FLUX ControlNet~\cite{zhang2023adding} conditioned on the decoder output. Neither model was trained on our codec outputs; instead, a generic version trained on common image corruptions (blur, JPEG, noise) was used. This approach yields modest perceptual gains (+0.5 dB DISTS) but significantly improves realism by restoring textures and fine details (Figure~\ref{fig:gen_decoder}).

\begin{table*}[htbp]
\centering
\begin{tabular}{@{}l rrrrr rrrrr r@{}}
\toprule
& \multicolumn{5}{c}{\textbf{Kodak} ($512\times768$ and $768\times512$)} & \multicolumn{6}{c}{\textbf{ImageNet-1k ($224\times224$)}} \\
\cmidrule(lr){2-6} \cmidrule(lr){7-12}
\textbf{Model} & \textbf{bpp} & \textbf{PSNR} & \textbf{LPIPS} & \textbf{DISTS} & \textbf{SSIM} & \textbf{bpp} & \textbf{PSNR} & \textbf{LPIPS} & \textbf{DISTS} & \textbf{SSIM} & \textbf{Acc} \\
\midrule
LiVeAction f16c12 (half res.)        & 0.0428 & 24.2336 & 2.7625 &  6.2970 & 0.7194 & 0.1112 & 23.647 & 3.5264 & 5.4999 & 0.6257 & 0.6277 \\
Cosmos di16$\times$16                & 0.0625 & 21.7743 & 5.3784 & 10.619  & 0.6449 & 0.0625 & 19.710 & 5.3340 & 8.4825 & 0.4882 & 0.7938 \\
AVIF (q=10)                          & 0.0916 & 26.5260 & 3.1129 &  6.9885 & 0.8080 & 0.3373 & 24.830 & 3.7429 & 5.7411 & 0.6982 & 0.7299 \\
JPEG 2000 (CR 250:1)                 & 0.0957 & 24.3492 & 2.4231 &  5.7073 & 0.6842 & 0.0978 & 20.450 & 2.0800 & 4.0033 & 0.4794 & 0.2807 \\
LiVeAction f16c12                    & 0.1496 & 27.0384 & 4.2153 &  8.8654 & 0.8600 & 0.1943 & 25.402 & 4.4948 & 6.3726 & 0.7122 & 0.7981 \\
Balle 2018 Hyperprior~\cite{balle2018variational}     & 0.2110 & 27.2377 & 3.9050 &  7.9973 & 0.8258 & 0.3338 & 28.188 & 5.3061 & 6.8009 & 0.8116 & 0.8360 \\
AVIF (q=25)                          & 0.2316 & 29.5176 & 4.4152 &  9.1244 & 0.9007 & 0.5281 & 27.789 & 5.3463 & 7.1298 & 0.8106 & 0.8262 \\
JPEG 2000 (CR 100:1)                 & 0.2393 & 26.4546 & 3.1854 &  7.1317 & 0.7829 & 0.2402 & 23.404 & 3.1483 & 5.0364 & 0.6014 & 0.6767 \\
Cosmos di8$\times$8                  & 0.2500 & 25.9193 & 7.7112 & 13.265  & 0.8558 & 0.2500 & 24.334 & 8.0240 & 10.808 & 0.7036 & 0.8708 \\
LiVeAction f16c48 ($\lambda=0.1$)    & 0.3992 & 29.8397 & 5.6911 & 12.073  & 0.9303 & 0.4868 & 27.978 & 6.1277 & 7.9186 & 0.8206 & 0.8463 \\
WaLLoC f8c12                         & 0.6171 & 30.5576 & 6.5138 & 13.244  & 0.9501 & 0.7720 & 29.102 & 7.0447 & 8.5500 & 0.8529 & 0.8569 \\
LiVeAction f16c48                    & 0.6606 & 31.1669 & 6.5692 & 13.670  & 0.9571 & 0.8072 & 29.501 & 7.1563 & 8.6339 & 0.8568 & 0.8560 \\
AVIF (q=50)                          & 0.6838 & 34.4449 & 7.1089 & 12.992  & 0.9657 & 1.1147 & 32.299 & 8.6288 & 9.9068 & 0.9144 & 0.8679 \\
JPEG 2000 (CR 20:1)                  & 1.1984 & 32.0019 & 5.4387 & 11.365  & 0.9262 & 1.1952 & 29.517 & 6.1171 & 7.7432 & 0.8206 & 0.8550 \\
WaLLoC f8c48                         & 2.5436 & 37.3370 & 11.674 & 18.294  & 0.9873 & 3.0067 & 35.107 & 12.120 & 13.063 & 0.9529 & 0.8838 \\
LiVeAction f16c192                   & 3.9126 & 40.2877 & 15.290 & 25.534  & 0.9962 & 4.2843 & 36.595 & 14.870 & 15.568 & 0.9640 & 0.8851 \\

\bottomrule
\end{tabular}
\caption{Additional results and baselines for RGB images. For ImageNet, the reported accuracy is the top-1 accuracy for 1000-way classification. The accuracy without lossy compression is 0.8979.}
\label{tab:image-compression}
\end{table*}

\section{Conclusion and Future Work}
We introduced LiVeAction, a neural codec design that establishes a new performance frontier and increases accessibility of learned compression for new types of signals and sensors. By improving signal-ingestion efficiency, LiVeAction lowers power and bandwidth demands while maintaining quality, enabling new mobile and remote sensing applications. Future work will explore variable-rate training and joint optimization with downstream ML tasks to better align compression with inference accuracy.

\bibliographystyle{IEEEbib}
\bibliography{refs}

\end{document}